\documentclass[conference]{IEEEtran}
\IEEEoverridecommandlockouts
\usepackage[dvipsnames]{xcolor}
\usepackage{mathtools}
\usepackage{amssymb}
\usepackage{graphicx}
\usepackage{cite}
\usepackage{booktabs}
\usepackage{dsfont}
\usepackage{tikz}
\usepackage{pgfplots}
\usepackage{array}
\usepackage{hyperref}
\usepackage{cleveref}
\usetikzlibrary{
    patterns,
    arrows.meta,
    positioning,
    fit,
    calc,
    shapes.geometric,
    shapes.arrows,
    shapes.symbols,
    shadows,
    shadings
}
\pgfplotsset{compat=1.18}
\definecolor{agentfill}{RGB}{216,231,241}
\definecolor{agentinput}{RGB}{218,220,240}
\definecolor{agenthidden}{RGB}{232,214,214}
\definecolor{agentoutput}{RGB}{211,230,213}

\definecolor{tacticalfill}{RGB}{230,238,230}

\definecolor{inferline}{RGB}{225,45,45}
\definecolor{planline}{RGB}{35,79,150}
\definecolor{execline}{RGB}{119,67,145}

\definecolor{calibblue}{RGB}{52,105,154}
\definecolor{recoverred}{RGB}{178,54,54}
\title{Hierarchical Agentic Incident Response with Digital-Twin-Validated Attack Inference}
\author{Yiran Gao\textsuperscript{1}, Juntao Chen\textsuperscript{2}, Tao Li\textsuperscript{1,\textdagger}\\
\textsuperscript{1}Department of Systems Engineering, City University of Hong Kong, \\
Hong Kong SAR, 99977, Emails: gaoyiran525@gmail.com, li.tao@cityu.edu.hk\\
\textsuperscript{2}Department of Computer and Information Sciences, Fordham University,\\
New York, NY, 10023, USA, Email: jchen504@fordham.edu

\thanks{Yiran Gao and Tao Li are supported by CityU internal grant No. 9610792. \textdagger Corresponding author: Tao Li.}

\vspace{-0.05in}
}

\begin{document}

\maketitle

\begin{abstract}
Network incident response remains slow and labor-intensive as the defender must infer multi-stage attacks from partial observations and translate recovery decisions into reliable system commands. Decision-theoretic planners provide principled optimization but typically rely on abstract states and predefined actions, while large language model (LLM) agents can reason over operational context but may hallucinate attacks and responses. Toward automating response planning, we present a hierarchical agentic response framework that integrates LLM-based attack inference, rollout planning, and digital-twin validation. A fine-tuned LLM infers the attack progression and affected hosts from security alerts and system measurements. An emulated network digital twin replays the inferred attack and returns discrepancies between predicted and observed effects to calibrate the inference. A separately fine-tuned planning agent uses the rollout planning method to prioritize affected components at the tactical layer. At the operational layer, the planning agent proposes high-level recovery actions, and an execution agent translates selected actions into recovery and verification commands that are validated in the digital twin. We evaluate the framework on a 33-component enterprise-network testbed under three multi-stage attack scenarios. The results show that our framework outperforms frontier-LLM baselines in recovery success rate by 18--31\%.
\end{abstract}

\section{Introduction}
Network incident response encompasses the analysis, containment, eradication, and recovery activities that follow an intrusion. Current practice relies heavily on human operators, who must interpret fragmented evidence and translate decisions into specific operations. These labor-intensive tasks are difficult to scale amid the persistent shortage of cybersecurity professionals.  Autonomous cyber defense (ACD) \cite{burnap25acd-survey} has therefore emerged as a promising approach in which security analysis and response are delegated to intelligent agents.

Decision-theoretic approaches to ACD use control and optimization, game theory, and reinforcement learning (RL) for response planning \cite{li2025agentic}.
They provide principled security decision-making through abstract models, including Markov decision processes (MDPs) and security games \cite{tao24ddztd}. However, they typically assume access to abstract security states and predefined defensive actions and are not designed to reconstruct concrete multi-stage attacks from partial observations or translate tactical decisions into executable recovery actions.

LLM agents offer a promising way to close this gap because they can interpret system context and security observations and generate system-specific response guidance and commands \cite{guo25ircopilot}. Recent works use prompt-based reasoning or multi-agent orchestration to select defensive actions and decompose incident-response tasks \cite{hamoun25llm-acd,guo25ircopilot}. Although these prior works include planning components, they generally rely on prompted reasoning rather than decision-theoretic sequential optimization. Consequently, their outputs can be sensitive to prompt design and available context, while hallucination remains an important concern \cite{kim26ndss}.

\begin{figure*}[t]
    \centering
    \resizebox{0.90\textwidth}{!}{%
        % Pure TikZ source for Fig. 1.
% Put this file at tikz/hierarchical_agentic_response.tex and include it
% inside a figure* environment from main.tex.
%
% Required in the main.tex preamble:
% \usetikzlibrary{arrows.meta,positioning,calc,shapes.geometric,
%                  shapes.arrows,shapes.symbols,shadows,shadings}
% \definecolor{agentfill}{RGB}{216,231,241}
% \definecolor{agentinput}{RGB}{218,220,240}
% \definecolor{agenthidden}{RGB}{232,214,214}
% \definecolor{agentoutput}{RGB}{211,230,213}
% \definecolor{inferline}{RGB}{220,73,46}
% \definecolor{planline}{RGB}{40,94,165}
% \definecolor{execline}{RGB}{124,73,151}
% \definecolor{calibblue}{RGB}{52,105,154}
% \definecolor{recoverred}{RGB}{178,54,54}

% Shared orange for tactical/operational scale labels.
\definecolor{scalecolor}{RGB}{214,105,31}

% Load the reusable digital-twin network component.
\input{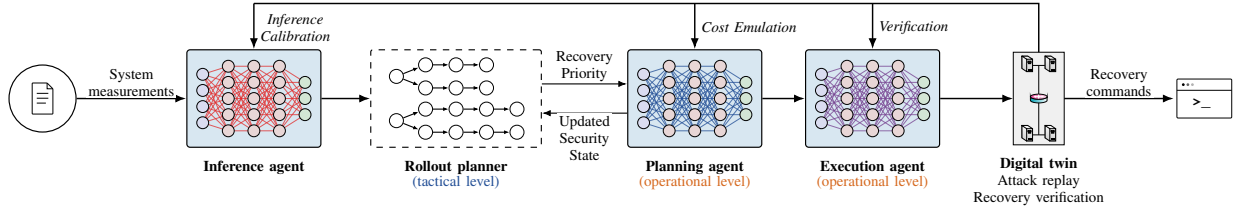}

% Compact fully connected 4-5-5-5-3 network.
% #1: name of the agent box; #2: color of the connections.
\newcommand{\drawagentnetwork}[2]{%
  \begin{scope}[shift={(#1.center)},x=4.20mm,y=2.65mm]
    % Coordinates for the five layers.
    \foreach \Layer/\N in {1/4,2/5,3/5,4/5,5/3}{%
      \foreach \i [evaluate={\x=\Layer-3;
                              \y=(\N+1)/2-\i;}] in {1,...,\N}{%
        \coordinate (#1-N\Layer-\i) at (\x,\y);
      }
    }

    % Connections are drawn first so that the neurons remain legible.
    \foreach \Layer/\N/\PreviousLayer/\PreviousN in
      {2/5/1/4,3/5/2/5,4/5/3/5,5/3/4/5}{%
      \foreach \i in {1,...,\N}{%
        \foreach \j in {1,...,\PreviousN}{%
          \draw[draw=#2,line width=0.20pt,opacity=0.78]
            (#1-N\PreviousLayer-\j) -- (#1-N\Layer-\i);
        }
      }
    }

    % All three agents use the same neuron colors.
    \foreach \Layer/\N in {1/4,2/5,3/5,4/5,5/3}{%
      \ifnum\Layer=1
        \def\NeuronFill{agentinput}%
      \else
        \ifnum\Layer=5
          \def\NeuronFill{agentoutput}%
        \else
          \def\NeuronFill{agenthidden}%
        \fi
      \fi
      \foreach \i in {1,...,\N}{%
        \node[circle,draw=black,fill=\NeuronFill,
              line width=0.32pt,minimum size=1.95mm,inner sep=0pt]
          at (#1-N\Layer-\i) {};
      }
    }
  \end{scope}%
}

% Two candidate component-priority roots, each with two rollout paths.
\newcommand{\drawdecisionplanner}[1]{%
  \begin{scope}[shift={(#1.center)},x=4.35mm,y=3.95mm]
    \foreach \Row/\RowY/\ExtraNode in {top/0.92/0,mid/-0.92/1}{%
      % Create named nodes first, as in deployment2.tex.  Connecting named
      % nodes makes every arrow terminate at the circle boundary.
      \node[circle,draw=black,fill=white,line width=0.38pt,
            minimum size=2.30mm,inner sep=0pt]
        (root-\Row) at (-2.25,\RowY) {};
      \foreach \Step/\X in {1/-1.15,2/0.00,3/1.15}{%
        \node[circle,draw=black,fill=white,line width=0.38pt,
              minimum size=2.30mm,inner sep=0pt]
          (up\Step-\Row) at (\X,{\RowY+0.48}) {};
        \node[circle,draw=black,fill=white,line width=0.38pt,
              minimum size=2.30mm,inner sep=0pt]
          (dn\Step-\Row) at (\X,{\RowY-0.48}) {};
      }
      \ifnum\ExtraNode=1
        \node[circle,draw=black,fill=white,line width=0.38pt,
              minimum size=2.30mm,inner sep=0pt]
          (up4-\Row) at (2.30,{\RowY+0.48}) {};
        \node[circle,draw=black,fill=white,line width=0.38pt,
              minimum size=2.30mm,inner sep=0pt]
          (dn4-\Row) at (2.30,{\RowY-0.48}) {};
      \fi

      % Each transition is an explicit rollout arrow, matching deployment2.tex.
      \draw[-{Latex[length=0.9mm]},line width=0.18mm,
            rounded corners=2pt,draw=black]
        (root-\Row) to (up1-\Row);
      \draw[-{Latex[length=0.9mm]},line width=0.18mm,
            rounded corners=2pt,draw=black]
        (up1-\Row) to (up2-\Row);
      \draw[-{Latex[length=0.9mm]},line width=0.18mm,
            rounded corners=2pt,draw=black]
        (up2-\Row) to (up3-\Row);
      \draw[-{Latex[length=0.9mm]},line width=0.18mm,
            rounded corners=2pt,draw=black]
        (root-\Row) to (dn1-\Row);
      \draw[-{Latex[length=0.9mm]},line width=0.18mm,
            rounded corners=2pt,draw=black]
        (dn1-\Row) to (dn2-\Row);
      \draw[-{Latex[length=0.9mm]},line width=0.18mm,
            rounded corners=2pt,draw=black]
        (dn2-\Row) to (dn3-\Row);
      \ifnum\ExtraNode=1
        \draw[-{Latex[length=0.9mm]},line width=0.18mm,
              rounded corners=2pt,draw=black]
          (up3-\Row) to (up4-\Row);
        \draw[-{Latex[length=0.9mm]},line width=0.18mm,
              rounded corners=2pt,draw=black]
          (dn3-\Row) to (dn4-\Row);
      \fi
    }
  \end{scope}%
}

% Remove whitespace accumulated before the TikZ picture.
\loop\ifdim\lastskip>0pt\unskip\repeat%
\begin{tikzpicture}[
    font=\scriptsize,
    >=Latex,
    flow/.style={-{Latex[length=1.8mm]},line width=0.68pt},
    exchange/.style={-{Latex[length=1.45mm]},line width=0.52pt},
    replay/.style={-{Latex[length=1.8mm]},line width=0.68pt,
                   draw=black},
    feedback/.style={-{Latex[length=1.6mm]},line width=0.58pt,
                     dashed,draw=recoverred},
    agentbox/.style={draw,rounded corners=1.5pt,line width=0.55pt,
                    fill=agentfill,minimum width=22mm,
                    minimum height=16mm,inner sep=1.5pt},
    labelbox/.style={fill=white,inner sep=0.8pt,align=center},
    modulelabel/.style={align=center,font=\scriptsize}
]

% Logs and system context input.
\node[
  circle,
  draw=black,
  fill=white,
  line width=0.55pt,
  minimum size=11mm,
  inner sep=0pt
] (logs) {};

% Folded document symbol, drawn entirely in TikZ.
\begin{scope}[shift={(logs.center)},x=1mm,y=1mm]
  \draw[line width=0.42pt,rounded corners=0.35pt]
    (-1.70,-2.55) -- (-1.70,2.55) -- (0.75,2.55)
    -- (1.70,1.60) -- (1.70,-2.55) -- cycle;
  \draw[line width=0.42pt]
    (0.75,2.55) -- (0.75,1.60) -- (1.70,1.60);
  \draw[line width=0.36pt] (-0.95,0.85) -- (0.95,0.85);
  \draw[line width=0.36pt] (-0.95,0.05) -- (0.95,0.05);
  \draw[line width=0.36pt] (-0.95,-0.75) -- (0.55,-0.75);
\end{scope}

\node[agentbox,right=18mm of logs] (infer) {};
\drawagentnetwork{infer}{inferline}
\node[modulelabel,below=0.9mm of infer] (inferlabel)
  {\textbf{Inference agent}};

\node[draw,dashed,rounded corners=1pt,line width=0.52pt,
      minimum width=28mm,minimum height=16mm,fill=white,
      right=8mm of infer] (tactical) {};
\drawdecisionplanner{tactical}
\node[modulelabel,below=0.9mm of tactical] (tacticallabel)
  {\textbf{Rollout planner}\\[-1pt]
   \textcolor{planline}{(tactical level)}};

\node[agentbox,right=14mm of tactical] (planning) {};
\drawagentnetwork{planning}{planline}
\node[modulelabel,below=0.9mm of planning] (planninglabel)
  {\textbf{Planning agent}\\[-1pt]
   \textcolor{scalecolor}{(operational level)}};

\node[agentbox,right=7mm of planning] (execution) {};
\drawagentnetwork{execution}{execline}
\node[modulelabel,below=0.9mm of execution] (executionlabel)
  {\textbf{Execution agent}\\[-1pt]
   \textcolor{scalecolor}{(operational level)}};

\node[
  inner sep=0pt,
  outer sep=0pt,
  minimum width=8.36mm,
  minimum height=15.20mm,
  right=12mm of execution
] (dticon) {};
\drawnetworkedsystemcomponentat{dticon.center}
\node[modulelabel,below=0.9mm of dticon] (dtlabel)
  {\textbf{Digital twin}\\Attack replay\\Recovery verification};

% Light terminal icon for the verified response commands.
\node[
  inner sep=0pt,
  outer sep=0pt,
  minimum width=10mm,
  minimum height=7mm,
  right=18mm of dticon
] (commandicon) {};

\begin{scope}[shift={(commandicon.center)},x=1mm,y=1mm]
  \draw[
    rounded corners=0.55pt,
    line width=0.45pt,
    fill=white
  ] (-4.50,-3.20) rectangle (4.50,3.20);
  \draw[line width=0.38pt]
    (-4.50,1.35) -- (4.50,1.35);
  \fill[black]    (-3.45,2.28) circle[radius=0.23];
  \fill[black!60] (-2.70,2.28) circle[radius=0.23];
  \fill[black!30] (-1.95,2.28) circle[radius=0.23];
  \node[font=\ttfamily\bfseries\scriptsize]
    at (-0.45,-0.75) {>\_};
\end{scope}

% Forward path.
\draw[flow] (logs) --
  node[labelbox,above,align=center] {System\\measurements}
  (infer);
\draw[flow] (infer) -- (tactical);
\draw[flow] (planning) -- (execution);
\draw[flow] (execution) -- (dticon);
\draw[flow] (dticon) --
  node[labelbox,above,align=center] {Recovery\\commands}
  (commandicon);

% Tactical-to-operational exchange.
\draw[exchange] ([yshift=2.4mm]tactical.east)
  -- node[labelbox,above,align=center] {Recovery\\Priority}
  ([yshift=2.4mm]planning.west);
\draw[exchange] ([yshift=-2.4mm]planning.west)
  -- node[labelbox,below,align=center] {Updated\\Security\\State}
  ([yshift=-2.4mm]tactical.east);

% Unified feedback bus originating from the digital twin.
\coordinate (feedbackBusDT) at ([yshift=8mm]dticon.north);
\coordinate (feedbackBusExecution) at (feedbackBusDT -| execution.north);
\coordinate (feedbackBusPlanning) at (feedbackBusDT -| planning.north);
\coordinate (feedbackBusInference) at (feedbackBusDT -| infer.north);
\draw[line width=0.68pt,draw=black]
  (dticon.north) -- (feedbackBusDT) -- (feedbackBusInference);

\draw[replay] (feedbackBusExecution)
  -- node[labelbox,pos=0.52,right=0.7mm] {\textit{Verification}}
  (execution.north);
\draw[replay] (feedbackBusPlanning)
  -- node[labelbox,pos=0.52,right=0.7mm] {\textit{Cost Emulation}}
  (planning.north);
\draw[replay] (feedbackBusInference)
  -- node[labelbox,pos=0.52,right=0.7mm,align=center]
     {\textit{Inference}\\\textit{Calibration}}
  (infer.north);

\end{tikzpicture}%
    }
    \vspace{-3mm}
    \caption{Overview of the hierarchical agentic incident-response framework.}
    \label{fig:hierarchical-agentic-response}
    \vspace{-4mm}
\end{figure*}

To bridge attack inference, tactical recovery planning, and operational execution, we develop a hierarchical response framework that integrates LLM agents, rollout planning, and digital-twin emulation. As shown in Fig.~\ref{fig:hierarchical-agentic-response}, a fine-tuned inference agent first reconstructs the sequence of core attack actions and their target hosts from system context and IDS observations. The digital twin then replays the inferred attack sequence and returns discrepancies between its observable effects and the incident observations to calibrate the inference. At the tactical layer, a lookahead rollout planner \cite{tao23cola} uses an abstract recovery-state model to prioritize affected components. At the operational layer, a fine-tuned planning agent proposes candidate high-level recovery actions for the prioritized component, and an execution agent translates each action into recovery and verification commands. Candidate actions are selected using Monte Carlo estimates of digital-twin-verified recovery costs.

We implement and evaluate the proposed framework on a containerized enterprise-network testbed comprising 33 interconnected components across multiple subnets. The environment includes multi-homed hosts, heterogeneous services with diverse vulnerabilities, IDS monitoring points, and attack paths requiring lateral movement across network segments. We consider three multi-stage attack scenarios with increasing attacker capability and attack-path complexity. Across these scenarios, our framework outperforms frontier-LLM baselines in recovery success rate by 18--31 percentage points.

Our contributions are summarized as follows.
\begin{itemize}
    \item We propose a hierarchical incident-response framework that integrates attack inference with tactical and operational recovery planning. At the tactical layer, decision-theoretic rollout prioritizes affected components across the network. At the operational layer, LLM agents propose component-specific recovery actions and translate the selected actions into executable commands.

    \item We ground both attack inference and recovery planning in a network digital twin. Inferred attack sequences are replayed and calibrated using discrepancies in their observable effects, while generated recovery commands undergo emulation-based verification.

    \item We implement the framework using locally deployed lightweight LLM agents and evaluate it on a 33-component, segmented enterprise-network testbed under three multi-stage attack scenarios. The framework outperforms frontier-LLM baselines in recovery success rate by 18--31 percentage points.
\end{itemize}

\section{Related Work}
\noindent{\it\bfseries Decision-theoretic incident response.}
Incident response is naturally a sequential decision-making problem in which states represent the network security posture and actions represent defensive responses. Recent studies have investigated online, meta, and multi-agent learning for adaptive defense \cite{kim-tao25col,kim-tao25quantization,tao23ztd,tao24col}. RL further enables data-driven planning when an explicit transition model is unavailable \cite{li2025agentic}. However, training and evaluating RL agents typically require a sufficiently faithful simulator or testbed, since extensive exploration on an operational network is unsafe and costly.

\noindent{\it\bfseries LLM-based incident response.}
LLMs offer a more operational alternative to decision-theoretic response methods by interpreting security observations and generating context-specific recovery actions. Existing approaches include prompt-based LLM orchestration \cite{hamoun25llm-acd,guo25ircopilot}, which typically does not perform explicit sequential optimization and can be sensitive to prompt design. Recent work combines decision-theoretic planning with LLM generation by using rollout or Monte Carlo tree search \cite{kim26ndss,yiran26e2e-ir}.

\noindent{\it\bfseries Cyber digital twins.}
Digital twins originated as digital replicas of physical systems that operate alongside their real-world counterparts to support situational awareness and decision making \cite{grieves2016digital,tao25dt-pirl,tao25dima}. In cybersecurity, digital twins can support high-fidelity emulation that reproduces operational system behavior in a controlled environment, enabling forensic analysis, vulnerability assessment, and the execution and validation of security operations before deployment \cite{repetto26cdt}.

\noindent{\it\bfseries Novelty of our approach.}
Unlike methods that assume the incident and affected assets are known, our framework uses a fine-tuned LLM to infer core attack actions and target hosts from system context and IDS observations and executes the inferred attack in a digital twin, thereby calibrating the inference against observable compromise effects. Compared with the most closely related approaches \cite{kim26ndss,yiran26e2e-ir}, our framework integrates decision-theoretic tactical planning, LLM-generated operational recovery actions, and digital-twin validation in a single hierarchy. This design combines abstract-model rollout for efficient tactical planning with digital-twin emulation for attack replay and operational recovery verification.

\section{Formalizing Incident Response Planning }
\label{sec:system_model}
We formulate incident response at two levels: (i) tactical planning, which prioritizes affected components for recovery, and (ii) operational planning, which selects recovery actions for a chosen component.
For cross-level planning, we consider a factorized partially observable Markov decision process, in which the system state factors into network-level security posture and component-level recovery states.

\noindent\textit{\textbf{System States, Observations, and Beliefs.}}
We represent the affected system by a graph $\mathcal{G}=(\mathcal{V},\mathcal{E})$, where $\mathcal{V}=\{1,\ldots,N\}$ is the set of system components and $\mathcal{E}$ captures their communication or dependency relationships.
The system operator, henceforth the \emph{defender}, monitors these components through intrusion-detection alerts, logs, and service measurements, which are referred to as partial observations. 

We define the global security state as an $N$-dimensional Boolean vector
\(
    g_t=(g_t^1,\ldots,g_t^N), g_t^k\in\{0,1\},
\)
where $g_t^k=1$ indicates that component $k$ is compromised, while $g_t^k=0$ indicates that it is safe.
Each component $k$ also has a local recovery state
\(
    \ell_t^k=(\ell_t^{k(c)},\ell_t^{k(a)},\ell_t^{k(p)},
    \ell_t^{k(e)},\ell_t^{k(h)},\ell_t^{k(r)})\in\{0,1\}^6,
\)
whose entries record completion of the six response stages \cite{d3fend}: 1) attack \textbf{containment}, 2) attack \textbf{assessment}, 3) forensic \textbf{preservation}, 4) attack \textbf{eviction}, 5) network \textbf{hardening}, and 6) service \textbf{restoration}. 
For example, $\ell_t^{k(c)}=1$ means that the attack on component $k$ has been contained.
The terminal local state is $\ell_{\mathrm{R}}=(1,1,1,1,1,1)$.
For a component that is initially safe, we set its local state to $\ell_{\mathrm{R}}$.
The joint system state across the two levels is
$s_t=((g_t^k,\ell_t^k)_{k\in[N]})$, $[N]\coloneqq\{1,\ldots, N\}$.

While the true system state remains hidden from the defender, the defender receives an observation $o_t$ that includes IDS alerts and system measurements and is correlated with $s_t$. Based on the historical observations $o_{0:t}$, the defender forms a state belief
$b_t(s)=\Pr(s_t=s\mid o_{0:t})$. Following the state factorization across the global and local levels, we represent $b_t$ by a global belief $b_t^{g}\in[0,1]^N$ and local beliefs $b_t^{k}\in[0,1]^6$ for $k\in[N]$.
Here, $b_t^{g}(k)$ is the probability that component $k$ is safe, and $b_t^{k}(i)$ is the probability that its $i$th response stage is complete. These beliefs are based on the assumption that recovery progress across components is independent, thereby reducing the dimensionality of the belief space. 

\noindent\textit{\textbf{Hierarchical Action and Cost.}} A response action is the pair $a_t=(a_t^g,a_t^\ell)$.
The tactical action $a_t^g$ is a permutation of $[N]$ that specifies the response priority of each component, where $a_t^g(k)$ is the priority rank of component $k$.
For a global state $g$, let $\mathcal{U}(g)=\{j\in[N]:g^j=1\}$ denote the set of compromised components that have not yet been recovered. Denote by $k$ the highest-priority component in $\mathcal{U}(g)$ under priority order $a^g$.
The operational action $a_t^\ell$ is the high-level recovery action applied to the selected component.

Let $\tau(\ell_t^k,a_t^\ell)$ denote the wall-clock time for completing $a_t^\ell$ on component $k$ when its local state is $\ell_t^k$. During this interval, component $k$ incurs its own execution time, while every other unrecovered component incurs the same amount of delay. We therefore define the unweighted aggregate recovery-delay cost as
\(
c(s_t,a_t)
=\left\lvert\mathcal{U}(g_t)\right\rvert\tau(\ell_t^k,a_t^\ell).
\)

\noindent\textit{\textbf{System Transition.}} Let $\theta\in\Theta$ denote the attack-action sequence executed during an incident, where $\Theta$ denotes the set of admissible attack sequences. Specifically, define the actual attack sequence as 
\(
\theta=((u_1,v_1),\ldots,(u_{d},v_{d})),
\)
where $d$ is the number of attack steps, $u_j$ is the core attack action at step $j$, and $v_j\in\mathcal{V}$ is its target component.

The attack sequence influences the component-level recovery dynamics. Let $\ell_t=(\ell_t^1,\ldots,\ell_t^N)$ collect the local recovery states, and let $k$ be the component selected by the tactical action. Because the operational action is applied only to component $k$, the local transition is
\begin{equation*}
P_\theta^\ell(\ell_{t+1}\mid\ell_t,a_t)
=P_\theta^k(\ell_{t+1}^k\mid\ell_t^k,a_t^\ell)
\prod_{j\ne k}\mathds{1}\{\ell_{t+1}^j=\ell_t^j\},
\end{equation*}
where $P_\theta^k$ describes the progress of component $k$ through the response stages. The realized local states then drive the global transition:
$
P_\theta(s_{t+1}\mid s_t,a_t)
=P_\theta^\ell(\ell_{t+1}\mid\ell_t,a_t)
P^g(g_{t+1}\mid g_t,\ell_{t+1}).$
Since $g_t^j$ records only whether component $j$ remains compromised, its update is deterministic:
\begin{equation*}
P^g(g_{t+1}\mid g_t,\ell_{t+1})
=\prod_{j=1}^N
\mathds{1}\!\left\{
g_{t+1}^j=g_t^j\mathds{1}\{\ell_{t+1}^j\ne\ell_{\mathrm{R}}\}
\right\}.
\end{equation*}
Thus, a safe component remains safe, whereas a compromised component becomes safe exactly when its local state reaches complete recovery. The complete-recovery state is $s_{\mathrm{R}}=((0,\ell_{\mathrm{R}})_{j\in[N]})$ and is absorbing: $P_\theta(s_{\mathrm{R}}\mid s_{\mathrm{R}},a)=1$ for every response action.

\noindent\textit{\textbf{Planning objective.}}
A response policy $\pi$ maps the current belief to a hierarchical response action, i.e., $a_t\sim\pi(\cdot\mid b_t)$.
The defender seeks a response policy $\pi$ that minimizes the expected cumulative recovery cost over a finite horizon $H$ under the transition model induced by the inferred attack sequence:
\begin{equation}
\min_{\pi}\;\mathbb{E}_{P_{\theta},\pi}\!\left[\sum_{t=0}^{H}c(s_t,a_t)\right].
\label{eq:obj}
\end{equation}
The expectation accounts for the uncertainty in both the partially observed recovery state and its evolution under the response policy.
The next section develops a hierarchical planning method that approximates this objective across the tactical and operational levels.

\section{Hierarchical Planning in Agentic Response}

\label{sec:planning}
Our method consists of an offline fine-tuning stage and an online hierarchical response stage. 
In the offline stage, we fine-tune two variants of a lightweight local model, DeepSeek-R1-Distill-Qwen-14B \cite{deepseek-r1}, on three instruction-response tasks. The first variant, referred to as the \textit{inference agent}, is trained on incident examples and aims to infer the attack sequence $\theta$. The second variant, called the \textit{planning agent}, is tasked with generating beliefs and actions.
In the online stage, the inference agent reconstructs and calibrates the attack sequence, after which tactical rollout prioritizes the affected components and the planning agent generates high-level recovery actions.
Finally, an \textit{execution agent}, powered by Qwen3.6-27B, translates selected high-level recovery actions into executable recovery and verification commands, which are validated in the digital twin.

\subsection{Offline Fine-Tuning}
We fine-tune the DeepSeek model, denoted by $\Phi_w$, using low-rank adaptation (LoRA) on an open-source corpus of approximately $68{,}000$ security incidents and corresponding responses, as in the previous work \cite{gao26multiscale}.
Let $\mathcal{D}=\{(\mathbf{x}^{i},\mathbf{y}^{i})\}_{i=1}^{K}$ denote the resulting instruction--response dataset, where $\mathbf{x}^{i}$ contains the task instruction and incident context and $\mathbf{y}^{i}=(y_1^i,\ldots,y_{L_i}^i)$ is the target output sequence.
We optimize the trainable model parameters $w$ using the autoregressive cross-entropy objective
\begin{equation}
\mathcal{L}(w)=-\frac{1}{B}\sum_{i=1}^{B}\sum_{n=1}^{L_i}\log\Phi_w\!\left(y_n^i\mid\mathbf{x}^{i},y_{1:n-1}^{i}\right),
\label{eq:fine-tuning-loss}
\end{equation}
where $B$ is the mini-batch size, $L_i$ is the number of target tokens in training example $i$, and $\Phi_w(y_n^i\mid\mathbf{x}^i,y_{1:n-1}^i)$ denotes the model's conditional probability of target token $y_n^i$.

Under task-specific instructions, the fine-tuned model performs three functions for use in the online procedure. 1) For the attack inference tasks (inference agent), the task instruction $\mathbf{x}$ includes the incident description and IDS alerts, while the response $
\mathbf{y}$ contains the corresponding MITRE ATT\&CK tactics and techniques, as well as attack sequences. 2) For state prediction (planning agent), the task instruction $\mathbf{x}$ remains the same, while the response contains the labeled state $s_t$ and its supporting evidence summary $m_t$. During deployment, we sample multiple state predictions and use their empirical frequencies to construct the belief $b_t$. 3) For action generation (planning agent), the instruction $\mathbf{x}$ is further augmented with the belief $b_t$, evidence summary $m_t$, and previous response action $a_{t-1}$, while the response contains the next recovery action recorded in the training dataset. In summary, we obtain two model weights, $w_I$ and $w_P$, for the two agents, respectively.

\subsection{Digital Twin-Based Attack Inference and Calibration}
The online stage unfolds as follows. The defender first invokes the inference agent, $\Phi_{w_I}$, to estimate the unknown attack sequence $\theta$ by inspecting the available partial observations $o_0$, including system context, IDS alerts, logs, and service measurements. The conjecture $\hat{\theta}$ is then executed in a network digital twin that replicates the affected system to produce an emulated observation $\hat o_0$. We deem $\hat o_0$ and $o_0$ consistent when they agree on the affected hosts, privilege and service states, backdoor indicators, IDS alerts, and attack-path reachability. If they are consistent, the inferred attack sequence is accepted for response planning. Otherwise, the observed discrepancies are returned to the inference agent, which revises the attack sequence before response planning begins. Calibration can repeat until the observations are consistent or a preset iteration limit is reached. In our experiments, we allow one revision because it already provides a satisfactory improvement in attack-inference accuracy, as shown in \Cref{tab:attack-action-frequency}.

% Let $o_0$ denote the defender-visible incident observation available before recovery, including system context, IDS alerts, logs, and service measurements.
% The true attack sequence $\theta^{\star}$ is unknown.
% The inference agent generates an inferred attack sequence
% \begin{equation}
% \hat{\theta}=\Phi_w(o_0).
% \label{eq:attack-inference}
% \end{equation}
% We execute $\hat{\theta}$ in the digital twin.
% For any executable attack sequence $\theta$, let $o_{\theta}$ be the defender-visible outcome produced by executing it in the digital twin:
% \begin{equation}
% (\tilde{s}_{\theta},o_{\theta})=\operatorname{DT\text{-}Attack}(\theta).
% \label{eq:dt-attack-observation}
% \end{equation}
% The replayed attack yields the predicted observation $\hat{o}\coloneqq o_{\hat{\theta}}$, which records observable effects such as compromised hosts, active sessions, privilege evidence, backdoor indicators, service states, IDS alerts, and attack-path reachability.
% If $o_0$ and $\hat{o}$ agree on the compromised hosts,  backdoor indicators, service states, IDS alerts, and attack-path reachability, the inferred sequence is accepted for recovery planning.
% Otherwise, the mismatched fields between $o_0$ and $\hat{o}$ are returned to the \textit{inference agent}, which revises the attack sequence before recovery begins.
% This grounds attack inference in digital-twin execution rather than textual agreement alone.

\subsection{Tactical Rollout Planning}
The inferred attack sequence identifies the affected components and induces an estimated global transition model ${P}_{\hat{\theta}}^g(g_{t+1}\mid g_t,a_t^g)$ for tactical planning.
At each tactical step, the priority order $a_t^g$ selects the highest-priority unrecovered component. The model assumes that this component is fully recovered through a sequence of operational actions before the next tactical decision, while the global states of the other components remain unchanged.
At recovery step $t$, the tactical planner constructs a candidate set $\mathcal{A}_t^g$ of component-priority permutations and evaluates which component should be recovered next.
For each candidate priority order $\tilde{a}^g$, let $k$ denote the highest-priority component in $\mathcal{U}(g)$. Let $\widehat{\tau}_t^k$ denote the estimated duration of recovering component $k$, computed from previously completed local recoveries when available and initialized using an offline average otherwise. Recovering $k$ delays every other unrecovered component by this duration. We therefore define the unweighted tactical delay cost as
\(
\widehat{c}^{g}(g,\tilde{a}^{g})
=\left\lvert\mathcal{U}(g)\setminus\{k\}\right\rvert
\widehat{\tau}_t^k.
\)
The planner samples $M_g$ possible global states $\{\hat{g}_t^i\}_{i=1}^{M_g}$ from the global belief $b_t^g$ and simulates each candidate for $H_g$ lookahead steps.
Its estimated rollout cost is given by
\begin{equation}
J_g(\tilde{a}^{g})=\frac{1}{M_g}\sum_{i=1}^{M_g}\sum_{h=0}^{H_g}\widehat{c}^{g}(\hat{g}_{t+h}^{i},\tilde{a}^{g}),
\label{eq:tactical-rollout-cost}
\end{equation}
where $\hat{g}_{t+h+1}^{i}\sim P_{\hat{\theta}}^g(\cdot\mid\hat{g}_{t+h}^{i},\tilde{a}^{g})$ for $h=0,\ldots,H_g-1$.
The selected tactical action is
$
a_t^g\in\arg\min_{\tilde{a}^{g}\in\mathcal{A}_t^g}J_g(\tilde{a}^{g}).
$
Since enumerating all $N!$ priority orders is intractable for large networks, the initial order ranks components by decreasing compromise probability $1-b_0^g(k)$. At subsequent recovery steps, $\mathcal{A}_t^g$ is restricted to the priority orders obtained by permuting up to the three highest-priority unrecovered components in the previous order.
Under the selected priority order $a_t^g$, let $k$ denote the highest-priority unrecovered component. This component is passed to operational planning.

\subsection{Operational Planning and Command Validation}
Let $k$ denote the component selected by the tactical priority order $a_t^g$.
Conditioned on the current belief $b_t$, its supporting evidence summary $m_t$, and the previous local action, the planning agent generates a set of candidate high-level recovery actions
$\mathcal{A}_t^{\ell}=\{\hat{a}_t^1,\ldots,\hat{a}_t^{N_{\ell}}\}$.
For each candidate $\hat{a}_t^i$, the model generates $M_{\ell}$ rollout trajectories $\{\zeta^{i,r}\}_{r=1}^{M_{\ell}}$, each beginning with $\hat{a}_t^i$ and continuing until the local terminal state $\ell_{\mathrm R}$ or the rollout-depth limit.

Every high-level action $\hat{a}$ in a trajectory is translated by the execution agent into a command plan $q(\hat{a})=(\mathcal{C}(\hat{a}),\mathcal{Q}(\hat{a}))$, where $\mathcal{C}(\hat{a})$ and $\mathcal{Q}(\hat{a})$ contain the recovery and verification commands, respectively.
The first validation stage checks whether the recovery and verification commands are compatible with the available containers, files, services, interfaces, and network paths.
The second stage executes the plan in the emulated digital twin and verifies whether it produces the intended recovery-state transition.
We define the verified operational cost as
\begin{equation*}
c_{\mathrm{DT}}(\hat{a})=
\begin{cases}
\operatorname{Time}_{\mathrm{DT}}(q(\hat{a})), & \text{if both validation stages pass},\\
\infty, & \text{otherwise},
\end{cases}
\end{equation*}
where $\operatorname{Time}_{\mathrm{DT}}$ is the wall-clock time required to execute the recovery and verification commands in the digital twin. Command generation, compatibility checking, and checkpoint restoration are measured separately as planning overhead.
All candidate rollouts begin from the same restored digital-twin checkpoint, ensuring that their costs are comparable, while actions within each rollout are executed sequentially. After all candidates are evaluated, the digital twin is restored once more to the pre-decision checkpoint before the selected command plan is executed.
A candidate is considered feasible only if every command plan in all of its sampled trajectories passes both validation stages; otherwise, its estimated cost is set to infinity.
The Monte Carlo estimate of the remaining operational recovery cost and the associated best response are, respectively, given by
\begin{equation*}
Q(b_t,\hat{a}_t^i)=\frac{1}{M_{\ell}}\sum_{r=1}^{M_{\ell}}\sum_{\hat{a}\in\zeta^{i,r}}c_{\mathrm{DT}}(\hat{a}), \quad a_t^{\ell}\in\arg\min_{\hat{a}_t^i\in\mathcal{A}_t^{\ell}}Q(b_t,\hat{a}_t^i).
\end{equation*}
If every candidate has infinite cost, no operational action is committed, and the planning attempt is declared unsuccessful. Otherwise, the command plan for the selected action $a_t^\ell$ is executed in the digital twin. The resulting recovery observation $o_{t+1}$, together with the current belief, evidence summary, and response action, is then returned to the planning agent. Multiple updated state predictions are sampled to construct $b_{t+1}$ from their empirical frequencies and update the supporting evidence summary $m_{t+1}$.
This generate--rollout--validate--execute cycle continues until the local state of component $k$ reaches $\ell_{\mathrm R}$ and the component is marked safe, after which tactical planning selects the next component.

\begin{table*}[!t]
\centering
\caption{Core attack actions in the three evaluation scenarios.}
\label{tab:attack-scenarios-summary}
\scriptsize
\setlength{\tabcolsep}{5pt}
\renewcommand{\arraystretch}{0.95}

\begin{tabular}{
  >{\raggedright\arraybackslash}p{0.14\textwidth}
  >{\raggedright\arraybackslash}p{0.76\textwidth}
}
\toprule
Scenario & Core attack actions \\
\midrule

Novice &
\texttt{TELNET\_SAME\_USER\_PASS\_DICTIONARY},
\texttt{FTP\_SAME\_USER\_PASS\_DICTIONARY},
\texttt{SHELLSHOCK\_EXPLOIT},
\texttt{SSH\_SAME\_USER\_PASS\_DICTIONARY},
\texttt{CVE\_2010\_0426\_PRIV\_ESC}
\\

Experienced &
\texttt{SAMBACRY\_EXPLOIT},
\texttt{SSH\_SAME\_USER\_PASS\_DICTIONARY},
\texttt{CVE\_2010\_0426\_PRIV\_ESC},
\texttt{DVWA\_SQL\_INJECTION},
\texttt{CVE\_2015\_1427\_EXPLOIT}
\\

Expert &
\texttt{SAMBACRY\_EXPLOIT (edge Samba)},
\texttt{DVWA\_SQL\_INJECTION},
\texttt{CVE\_2015\_1427\_EXPLOIT},
\texttt{SAMBACRY\_EXPLOIT (deep internal Samba)}
\\

\bottomrule
\end{tabular}
\vspace{-5mm}
\end{table*}

\section{Experimental Evaluation}
\label{sec:experiment}
We evaluate the end-to-end recovery performance of the proposed framework using recovery time and recovery rate.
The training-data construction, fine-tuning procedure, and standalone evaluation of the local DeepSeek-R1-Distill-Qwen-14B model are reported in our previous work \cite{gao26multiscale}; hence, we do not repeat the model-level evaluation here because of space limitations.

\subsection{Experimental Setup}
\begin{figure}[t]
    \centering
    \includegraphics[
        width=0.8\columnwidth,
        keepaspectratio
    ]{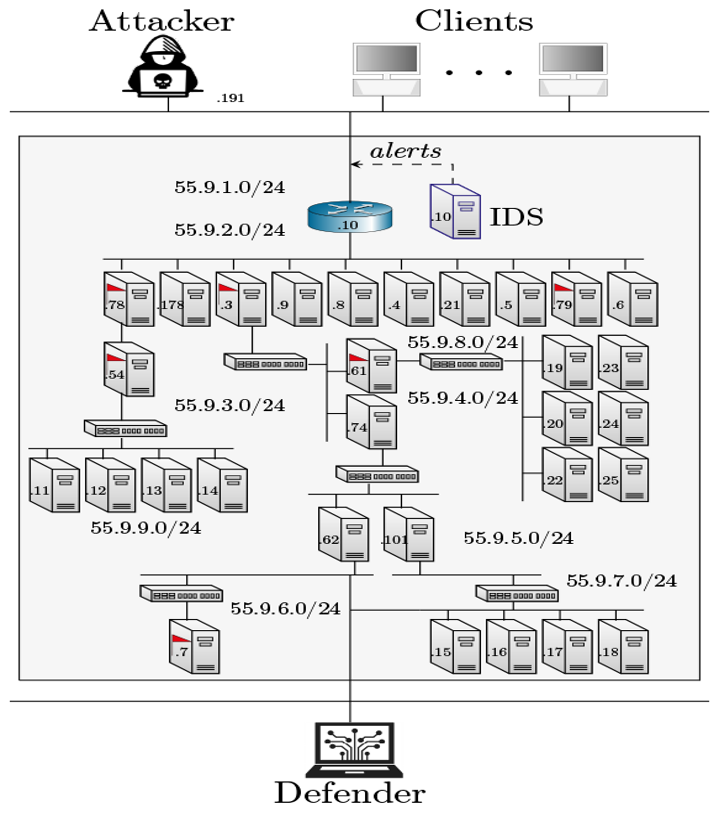}
    \caption{Enterprise-network digital-twin topology.}
    \label{fig:level9-topology}
    \vspace{-3mm}
\end{figure}

\noindent\textit{\textbf{Digital-twin environment.}}
We conduct the experiments on a containerized enterprise network digital twin comprising approximately 33 interconnected components distributed across multiple subnets, adapted from \cite{hammarcsle}.
The environment includes multi-homed hosts, heterogeneous vulnerable services, routing components, and IDS monitoring points.
This topology, as shown in Fig.~\ref{fig:level9-topology}, supports attack paths that cross network segments and therefore requires the response planner to reason about affected components and their network dependencies.

\noindent\textit{\textbf{Attack scenarios.}}
We consider three multi-stage attack scenarios, denoted \textit{novice}, \textit{experienced}, and \textit{expert}.
The scenarios represent increasing attacker capability and attack-path complexity, ranging from a comparatively short compromise path to attacks involving additional exploitation and lateral movement across multi-homed components. Table~\ref{tab:attack-scenarios-summary} summarizes the core attack actions in each scenario.

\noindent\textit{\textbf{Implementation.}}
The local inference model is DeepSeek-R1-Distill-Qwen-14B with the fine-tuned adapter described in \cite{gao26multiscale}.
We use the same four-dataset LoRA fine-tuning configuration as in \cite{gao26multiscale}.
At each operational decision step, the model proposes two candidate high-level recovery actions, with two Monte Carlo rollouts per candidate.
The {execution agent} translates semantic actions into component-specific recovery and verification commands, which are validated and executed in the digital twin.
All experiments are performed on a Google Cloud virtual-machine instance equipped with one NVIDIA A100 GPU with 40 GB of memory.

\subsection{Metrics and Baselines}

\noindent\textit{\textbf{Core attack-action accuracy.}}
We first evaluate whether the fine-tuned LLM can infer the core attack actions used in each multi-stage attack.
For each scenario, we generate 20 predictions from the same system description and IDS observations, normalize each into an attack-action set, and compare it with the reference set. Per-action accuracy is the fraction of predictions containing a given ground-truth action. 
Let True Positive (TP) denote correctly inferred attack actions, False Positive (FP) denote inferred actions that are not in the reference attack, and False Negative (FN) denote reference attack actions missed by the model.
Following the class-agnostic evaluation used in our prior work, we aggregate TP, FP, and FN over all repeated predictions before computing precision, recall, and F1.
They are computed as $\mathrm{Precision}=\mathrm{TP}/(\mathrm{TP}+\mathrm{FP})$, $\mathrm{Recall}=\mathrm{TP}/(\mathrm{TP}+\mathrm{FN})$, and $\mathrm{F1}=2\cdot\mathrm{Precision}\cdot\mathrm{Recall}/(\mathrm{Precision}+\mathrm{Recall})$.

\begin{table}[t]
\centering
\caption{Per-action attack inference accuracy across scenarios before and after calibration.}
\label{tab:attack-action-frequency}
\scriptsize
\setlength{\tabcolsep}{3pt}
\renewcommand{\arraystretch}{0.9}

\begin{tabular}{lcc}
\toprule
Novice & Pre-Calibration & Post-Calibration \\
\midrule
\texttt{SSH\_SAME\_USER\_PASS\_DICTIONARY}
    & 50\% & 55\% \\
\texttt{TELNET\_SAME\_USER\_PASS\_DICTIONARY}
    & 60\% & 70\% \\
\texttt{FTP\_SAME\_USER\_PASS\_DICTIONARY}
    & 25\% & 55\% \\
\texttt{SHELLSHOCK\_EXPLOIT}
    & 40\% & 45\% \\
\texttt{CVE\_2010\_0426\_PRIV\_ESC}
    & 75\% & 75\% \\

\midrule
Experienced & Pre-Calibration & Post-Calibration \\
\midrule
\texttt{SAMBACRY\_EXPLOIT}
    & 95\% & 95\% \\
\texttt{SSH\_SAME\_USER\_PASS\_DICTIONARY}
    & 90\% & 95\% \\
\texttt{CVE\_2010\_0426\_PRIV\_ESC}
    & 15\% & 65\% \\
\texttt{DVWA\_SQL\_INJECTION}
    & 90\% & 90\% \\
\texttt{CVE\_2015\_1427\_EXPLOIT}
    & 90\% & 90\% \\

\midrule
Expert & Pre-Calibration & Post-Calibration \\
\midrule
\texttt{SAMBACRY\_EXPLOIT} (edge Samba)
    & 100\% & 100\% \\
\texttt{DVWA\_SQL\_INJECTION}
    & 100\% & 100\% \\
\texttt{CVE\_2015\_1427\_EXPLOIT}
    & 100\% & 100\% \\
\texttt{SAMBACRY\_EXPLOIT} (deep internal Samba)
    & 100\% & 100\% \\

\bottomrule
\end{tabular}
\end{table}

\begin{table}[t]
\centering
\caption{Scenario-level precision, recall, and F1 scores for attack-action inference.}
\label{tab:attack-action-prf}
\scriptsize
\setlength{\tabcolsep}{3pt}
\renewcommand{\arraystretch}{0.9}
\begin{tabular}{cccc}
\toprule
Scenario & Novice & Experienced & Expert \\
\midrule
(P, R, F1; \%) &
(78.1, 50.0, 61.0) &
(98.7, 76.0, 85.9) &
(88.9, 100.0, 94.1) \\
\bottomrule
\end{tabular}
\vspace{-5mm}
\end{table}

\noindent\textit{\textbf{Recovery success and recovery time.}}
We then evaluate the recovery stage using recovery success rate, planning time, and operational execution time.
A trial is successful only if the selected recovery sequence reaches the terminal six-stage recovery state and the corresponding operational commands pass digital-twin validation.
Planning time includes high-level action generation, rollout evaluation, command generation and validation, and digital-twin state restoration. Operational execution time measures the execution and verification of the selected recovery commands.

\begin{table}[t]
    \centering
    \caption{Recovery planning and operational execution time.}
    \label{tab:level9-recovery-time}
    \scriptsize
    \setlength{\tabcolsep}{1.8pt}
    \renewcommand{\arraystretch}{0.90}
    \begin{tabular}{llccc}
        \toprule
        Scenario & Agent & Planning (min) & Execution (s) & Success \\
        \midrule
        Novice & Our agent & 79.8 $\pm$ 2.1 & 43.5 $\pm$ 4.8 & 90\% \\
        & GPT-5.5 & 28.5 $\pm$ 5.6 & 66.7 $\pm$ 12.4 & 72\% \\
        & Gemini-3.1-Pro & 34.4 $\pm$ 6.4 & 74.6 $\pm$ 14.2 & 68\% \\
        & Claude Opus 4.8 & 29.8 $\pm$ 5.8 & 69.4 $\pm$ 12.9 & 71\% \\
        \midrule
        Experienced & Our agent & 102.9 $\pm$ 7.7 & 137.9 $\pm$ 6.1 & 88\% \\
        & GPT-5.5 & 71.4 $\pm$ 8.5 & 142.8 $\pm$ 17.6 & 66\% \\
        & Gemini-3.1-Pro & 82.7 $\pm$ 9.6 & 153.8 $\pm$ 21.4 & 62\% \\
        & Claude Opus 4.8 & 73.9 $\pm$ 8.8 & 151.6 $\pm$ 18.7 & 67\% \\
        \midrule
        Expert & Our agent & 106.7 $\pm$ 5.6 & 144.4 $\pm$ 6.4 & 86\% \\
        & GPT-5.5 & 77.2 $\pm$ 6.3 & 148.6 $\pm$ 18.1 & 59\% \\
        & Gemini-3.1-Pro & 89.6 $\pm$ 7.8 & 165.4 $\pm$ 20.2 & 55\% \\
        & Claude Opus 4.8 & 80.4 $\pm$ 7.1 & 145.0 $\pm$ 17.8 & 60\% \\
        \bottomrule
    \end{tabular}
    \vspace{-5mm}
\end{table}

\subsection{Results Discussion}

For attack inference, precision exceeds 78\% across all three scenarios,
as shown in \Cref{tab:attack-action-prf}. Turning to recovery,
\Cref{tab:level9-recovery-time} shows that our method achieves recovery
success rates of 90\%, 88\%, and 86\% in the novice, experienced, and
expert scenarios, versus 72\%, 67\%, and 60\% for the strongest baselines,
yielding gains of 18, 21, and 26 percentage points.

Execution logs show that our planning agent typically commits six high-level actions per recovered component, corresponding to the six local recovery stages. In contrast, the baselines generate seven to fifteen actions, requiring more command-generation and execution rounds and potentially contributing to longer and more variable execution times.

\section{Conclusion}
We present a hierarchical agentic incident-response framework combining LLM-based attack inference, decision-theoretic planning, and digital-twin emulation. Digital-twin replay calibrates inferred attack progression, tactical rollout prioritizes affected components, and planning and execution agents generate recovery actions and commands that are validated in the twin. On a 33-component, multi-subnet testbed under three multi-stage attack scenarios, our framework improves recovery success over frontier-LLM baselines by 18--31 percentage points while achieving the shortest mean operational execution time in every scenario.
\bibliographystyle{IEEEtran}
\bibliography{ref}

\end{document}